\documentstyle[aps,prb,preprint]{revtex} 
\begin{document}
\draft






\title{Spin dynamics in hole-doped two-dimensional S=1/2 
Heisenberg antiferromagnets: $^{63}$Cu NQR relaxation in La$_{2-x}$Sr$_x$CuO$_4$ for $x\leq 0.04$}


\author{P. Carretta, F. Tedoldi, A. Rigamonti and F. Galli} 
\address{
Department of Physics "A. Volta", 
Unita' INFM di Pavia-Via Bassi, 6 - 27100-I Pavia, ITALY}


\author{F. Borsa} 
\address{
Department of Physics "A. Volta", 
Unita' INFM di Pavia-Via Bassi, 6 - 27100-I Pavia, ITALY
and Ames Lab.-USDOE and 
Dept. of Physics and Astronomy, ISU, Ames, Iowa 50011, USA}


\author{J. H. Cho and D. C. Johnston
}
\address{Ames Lab.-USDOE and 
Dept. of Physics and Astronomy, ISU, Ames, Iowa 50011, USA}


\date{\today}
\maketitle
\widetext
\begin{abstract}


The effects on the correlated Cu$^{2+}$ $S = 1/2$ spin dynamics 
in the paramagnetic phase of La$_{2-x}$Sr$_x$CuO$_4$ (for $x \lesssim 0.04$) due to 
the injection of holes are studied by means of $^{63}$Cu NQR spin-lattice relaxation time $T_1$  
measurements. The results are discussed in the framework of the connection between $T_1$ and 
the in-plane magnetic correlation length $\xi_{2D}(x,T)$. 
It is found that at high temperatures the system remains in the renormalized classical regime, 
with a spin stiffness constant $\rho_s(x)$  reduced by small doping to an extent larger than the
one due to Zn doping. For $x\gtrsim 0.02$ the effect of doping on $\rho_s(x)$ appears to
level off.
The values for $\rho_s(x)$ derived from $T_1$ for $T\gtrsim 500$ K are much larger
than the ones estimated from the temperature behavior of
sublattice magnetization in the  ordered phase ($T\leq T_N$).
It is argued that these features are consistent with
the hypothesis of formation of stripes of microsegregated holes.
\end{abstract}
\newpage
\narrowtext







Besides being the parent compound of high temperature superconductors, La$_2$CuO$_4$ 
 can be considered as a prototype for the investigation of quantum spin  
magnetism in planar square lattice Heisenberg antiferromagnets (2D-QHAF) 
\cite{bib1,bib2}. 
This compound shows, over a wide temperature range, $T_N\simeq 315 K < T < J \simeq 1500 K$, 
strong in-plane correlations without 3D long range order. Recent 
theories for 2D-QHAF \cite{bib3,bib4} 
predict that La$_2$CuO$_4$ above $T_N$ is in the renormalized 
classical (RC) regime, where the spin wave stiffness $\rho_s$ and the spin wave velocity 
$c_{sw}$ are renormalized by quantum fluctuations with respect to the correspondent 
classical mean field values. 
In the RC regime one expects for the in-plane magnetic correlation length 
\cite{bib4}
\begin{eqnarray}
\xi_{2D}/a={\hbar c_{sw}\over 16\pi k_B\rho_s}e^{2\pi\rho_s\over T}
\biggl[1- 0.5{T\over 2\pi\rho_s}\biggr]= \\ \nonumber
=0.493  e^{1.15 J/T}\biggl[1- 0.43{T\over J} + O({T\over J})^2\biggr] 
\end{eqnarray}
where  $a$ is the lattice constant.
The spin stiffness constant has been written 
$\rho_s = 1.15J/2\pi$, while $c_{sw}=1.18\sqrt{2}Jk_Ba/\hbar$, with $J$
in temperature units.


Recently \cite{bib5,bib6}, from a series of $^{63}$Cu and $^{139}$La NQR and $\mu$SR 
measurements in spin-doped La$_2$CuO$_4$, 
where Cu$^{2+}$ ions were substituted by Zn$^{2+}$(S = 0) ions, 
several aspects have 
been clarified:
i)  The pure La$_2$CuO$_4$ remains in the RC regime up to T$\simeq 900$ K.
ii) The values for $\xi_{2D}$ derived from $^{63}$Cu NQR relaxation rates are in close quantitative 
agreement with those derived from neutron scattering.
iii) Above $900$ K a possible crossover to a quantum critical
(QC) regime occurs.
iv) The effect of Zn-doping on $\xi_{2D}$ in the RC regime can be satisfactory described in terms 
of a dilution-like model, whereby the doping causes a decrease of the spin stiffness of the 
form $\rho_s(x)=\rho_s(0)[1-(2-x)x]$.
v) Up to $\simeq 800$ K there is no evidence of a crossover to a QC regime, even for Zn
doping levels up to $11$ \%. 
On the other hand, the effects of heterovalent (Sr$^{2+}$  for La$^{3+}$) substitutions, where itinerant
 holes coupled in singlet states with Cu$^{2+}$ spins perturb the 2D antiferromagnetic (AF) 
correlation, have not been entirely clarified yet. 
The problem of the effects of charge doping on the magnetic properties of the 
CuO$_2$ plane is of particular interest. In fact, besides causing 
the drastic decrease of the N\'eel temperature with doping ( $T_N\rightarrow 0$  
at $x=x_c=0.02$)
\cite{bib7,bib8} the role of the itinerant charges is relevant for  the 
mechanisms underlying high T$_c$ 
superconductivity, particularly in the light of the theoretical approaches based 
on the microscopic
segregation of the
itinerant holes along stripes \cite{bib2,Emery,Grilli}. 
In this report  $^{63}$Cu NQR spin-lattice relaxation measurements 
in the doping range just below and just above the critical concentration $x_c = 0.02$
are presented. It is argued how 
some evidence for the presence of dynamical charge separation in domain walls is
provided by the data.


Single phase samples of La$_{2-x}$Sr$_x$CuO$_4$ 
were prepared by solid state reaction and annealed in 1 bar O$_2$ 
\cite{bib7,bib8,bib9}. 
Oxygen exchange during the high temperature measurements was prevented by sealing the samples in 
pyrex ampoules. 
The $^{63}$Cu 
NQR frequency is in the range $33-33.5$ MHz depending on the composition of the sample. 
$^{63}$Cu relaxation rates have been obtained with a pulse 
spectrometer by monitoring the recovery of the echo amplitude after a sequence of pulses 
yielding equalization of the populations of the $\pm 1/2$ and $\pm 3/2$ $^{63}$Cu NQR levels. 
The relaxation rate  was extracted from the exponential recovery of the echo intensity. 
The experimental results for 2W as a function of temperature are shown in Fig. 1 for some 
samples with different concentrations $x$ of Sr dopant. Our results for 
$x=0.04$ , not shown in Fig.1, agree with the ones 
reported previously for the same concentration by other authors
\cite{bib10}. 
The enhancement of the relaxation rate as the temperature is lowered becomes 
less pronounced as the concentration of Sr increases. In all the samples investigated the 
NQR signal is lost well above the antiferromagnetic transition temperature $T_N$ 
due to the progressive decrease of the echo dephasing time $T_2$ on decreasing temperature 
\cite{bib5}. 


$^{63}$Cu nuclear spin-lattice relaxation in La$_2$CuO$_4$ 
is driven by the magnetic field fluctuations at the nuclear site and one can write 
\begin{equation}
1/T_1\equiv 2W = {\gamma^2\over 2}\int_{-\infty}^{+\infty}e^{-i\omega_Rt}<h_-(0)h_+(t)> dt
\end{equation}
where $\gamma$ is the $^{63}$Cu  gyromagnetic ratio, $\omega_R$ is the resonance frequency 
and  $h_{\pm}$ are the components of the fluctuating field transverse to the quantization axis 
(the $c$-axis, corresponding to the direction of the $V_{zz}$ electric
field gradient component). 
The hyperfine field at the Cu nucleus can be written $\vec h={\cal A}\vec S_o + \sum_{i=1}^4 B\vec S_i$,
where the on-site interaction constant is $A_{\perp}= 80$ kGauss, while the 
transferred hyperfine coupling constant is $B =83$ kGauss \cite{Rasse}. 
From the correlation function $<h_+(0)h_-(t)>$ in Eq. (3) 
one arrives at $2W=(\gamma^2/2)\sum_{\vec q}{\cal A}_{\vec q}^2\vert S_{\vec q}\vert^2 /\Gamma_{\vec q}$
, where $\vert S_{\vec q}\vert^2$ is the amplitude of the collective spin fluctuations, and 
$\Gamma_{\vec q}^{-1}$  
the corresponding  decay time. The coupling constant is 
${\cal A}_{\vec q}^2=[A_{\perp} - 2B(cos(q_xa)+cos(q_yb))]^2$, with $\vec q$ 
starting from $(\pi/a, \pi/a)$ \cite{Rasse}. 
Since in the temperature range of interest $\xi_{2D}\gg a$, 
scaling arguments for $\vert S_{\vec q}\vert^2$ and $\Gamma_{\vec q}$, 
or equivalently for the generalized susceptibility $\chi(\vec q,\omega)$, can be used. 
Then $\chi(\vec q,\omega)=\chi_o\xi^zf(q\xi,\omega/\xi^z)$, with
$\chi_o= S(S+1)/3k_BT$  
and z the dynamical scaling exponent, and one obtains \cite{bib5}
\begin{eqnarray}
2W=\gamma^2{S(S+1)\over 3}\epsilon({\xi_{2D}\over a})^{z+2}{\beta^2\sqrt{2\pi}\over\omega_e}({a^2
\over 4\pi^2})\times \\ \nonumber
\times \int_{BZ}d\vec q {[A_{\perp}-2B(cos(q_xa)+cos(q_yb))]^2\over (1+q^2\xi_{2D}^2)^2}
\end{eqnarray}
where $\omega_e$ is the 
Heisenberg exchange frequency describing the fluctuations in the limit of infinite temperature, 
$\epsilon=0.3$ takes into account the reduction of the 
amplitude due to quantum fluctuations \cite{bib3,bib11}
 and $\beta$ is a normalization factor preserving the total moment  sum rule. 
It is noted that a simple analytical form emphasizing the connections of $\xi_{2D}$ to 
$2W$ can be obtained by averaging over the Brillouin zone the form factor in square brackets 
of Eq. (3). In this case $2W\simeq 4.2\times 10^3 (\xi_{2D}/a)^z/[ln(q_m\xi_{2D})]^2 sec^{-1}$, 
where $q_m=2\sqrt{\pi}/a$ (see Ref. 5). 


The validity of Eq. (3) was found \cite{bib5,bib6}
 to extend up to temperatures $T =1000$ K  for La$_2$CuO$_4$,
where $\xi_{2D}/a\simeq 2$. Therefore
it could be argued that for low dimensional systems the validity of scaling arguments is not 
strictly limited to the range where $\xi_{2D}/a \gg 1$. In this respect it is worth pointing out 
 that the maximum in the magnetic susceptibility, which indicates the occurrence 
of substantial short range order, is  estimated for La$_2$CuO$_4$ around $1500$ K 
 \cite{bib2}, where $\xi_{2D}/a\simeq 1$. Therefore, one can safely use Eq. (3), with a
numerical integration over the Brillouin zone, to extract, from the experimental evaluation
of $T_1$, the temperature and doping dependence of $\xi_{2D}(x,T)$.
 The results are shown in Fig.2, for different 
 Sr concentrations.  The slope of the semilog plot of $\xi_{2D}$ vs. $1000/T$ 
decreases with increasing Sr concentration, reflecting the decrease of the spin stiffness 
$\rho_s(x)$ in Eq. (1).  For $x\lesssim 0.02$ the data can be fitted over 
all the explored temperature range  by using Eq.
(1), with $J= 1340$ K ($\pm 40$ K ) for $x = 0.012$ and $J= 1200$ K ($\pm 50$ K ) for $x = 0.018$  
respectively. The decrease of the spin stiffness with $x$, for $x\rightarrow 0$, is faster 
than expected for a diluted 2D AF, as Zn-doped La$_2$CuO$_4$ \cite{bib5} (see Fig. 3). 
The more pronounced decrease of the spin-stiffness
for Sr$^{2+}$ doping should be associated with the mobile nature of the holes which induce a 
larger damping in the spin excitations \cite{Acqua,CuOLi}. On the other hand it can be observed
that the decrease of
the spin-stiffness with $x$ is weaker than the one derived for holes randomly itinerating  
in the AF matrix, namely $\rho_s(x)\propto 1/x$ \cite{Acqua,CuOLi}. 
From this observation one is lead to conjecture that a reduced effective amount of holes
controls the spin stiffness for $x\gtrsim 0.015$
A reduction in the local carrier density could in principle
result from the microsegregation of the hole carriers along stripes (or segments
of stripes) 
which leave a hole depleted region in between them. Within this line of interpretation
one would conclude that 
La$_{2-x}$Sr$_x$CuO$_4$ remains in the RC regime with a spin-stiffness 
constant reduced with respect to the pure La$_2$CuO$_4$, up to the limit where  the correlation
length is smaller than the average distance $l$ between stripes. 


For large doping the increase in $T_2$ allows one to extend
$1/T_1$ measurements at lower temperatures.  For $T\gtrsim 550$ K even for $x\geq 0.02$
La$_{2-x}$Sr$_x$CuO$_4$ remains in the RC regime. However,
below about 550 K, for $x\gtrsim 0.02$, one has a flattening 
in the T dependence of $\xi_{2D}$.  
The tendency of $\xi_{2D}$  
towards saturation has been already observed through 
neutron scattering measurements
by Keimer at al. \cite{bib12} and described on the basis of the  
phenomenological expression $1/\xi(x,T)=1/\xi(x,0) + 1/\xi(0,T)$, 
where $\xi(0,T)$ is
given by Eq. (1) with $J = 1588$ K. 
The $x$ dependence of the 
correlation length for $T\rightarrow 0$  yields information on
the  
topology of the holes. If the holes are localized  close 
to the randomly distributed
Sr$^{2+}$ impurity ions, one expects
 $\xi(x,0)=a/\sqrt{x}= 3.8/\sqrt{x}$ \AA \cite{bib13}. 
On the other hand, if the correlation length is limited by the formation of 
domain walls where the mobile holes are segregated, then  $\xi(x,0)=a/nx$, 
where $n$ is the average distance between the holes along 
the domain walls in lattice units
\cite{bib14,bib15}. Although it is difficult to
distinguish the $x$ dependence of the form $1/x$ from the $1/\sqrt{x}$ one, it should be noticed 
that the estimated values of $\xi(x,0)$ do confirm the microsegregation. In fact,
the assumption of a random distribution of holes would imply values for $\xi(x,0)$ much
smaller than the ones experimentally measured. For example, for
$x=0.03$ one should have
$\xi(x,0)=a/\sqrt{x}\simeq 6a$, while the experimental value is 
$\xi(x,0)\simeq 20a$ (see Fig. 2). On the contrary, in the presence of stripes, with  $n=2$
(as found by Tranquada et al. \cite{bib15} in the  $x=1/8$ compound), one
obtains $\xi(x=0.03,0)\simeq 16a$, in close agreement with the experimental
finding.
  
  
Also these estimates appear to reinforce the idea that the holes segregate along 
domain walls, or stripes.
The occurrence of the stripes was originally proposed in order to justify the susceptibility 
\cite{bib7} and the $x$ dependence of the sublattice magnetization $M(x,0)$ in the limit
$T\rightarrow 0$, 
as derived from $^{139}$La NQR and $\mu$SR measurements \cite{bib14}.
A quantitative description of the effective spin stiffness $\rho_s(x)$ resulting from the presence
of the stripes below $T_N$ has been given by Castro Neto and Hone \cite{CNH} and more
recently by Van Duin and Zaanen \cite{VDZ}. These descriptions are based on the quantum non-linear
$\sigma$ model and assume an in-plane anisotropy for the superexchange constant $J$ in an 
effective Heisenberg hamiltonian. In this framework, for $T\rightarrow 0$ a pronounced
decrease of the spin stiffness with $x$ justifies the experimental behavior for $M(x,0)$
\cite{bib14}.


It is worth to compare the $x-$dependence of $\rho_s(x)$ derived in the high temperature range from
$^{63}$Cu NQR $T_1$ (Fig. 3) with the behavior for $\rho_s(x)$ expected at low temperature
on the basis of the picture by Castro Neto and Hone \cite{CNH}. In Fig. 3 the solid line shows the
$x-$dependence of the spin stiffness according to Eq. 7 in Ref. 21. The comparison with the data
for $\rho_s(x)$ derived from $^{63}$Cu NQR $T_1$ shows that the decrease of $\rho_s(x)$ from
$T_1$ is much smaller than the one derived from the staggered magnetization. 
The difference is related
to the different temperature regions 
probed by the two quantities, implying that two different 
regimes are present in the spin dynamics.
At high temperature, where $\xi_{2D}
<l$, the stripes are mobile, 
possibly corresponding to anti-phase boundaries \cite{CNH} and the 
spin excitations are the ones characteristic of the 2DQHAF with a
reduced number of holes. Accordingly, $\rho_s(x)$ is only  slightly reduced by 
doping. The characteristic fluctuation  time for the stripes
is much longer than $\omega_e^{-1}$ and in the life time of the spin excitations
the stripes appear as nearly static.
On the contrary, in the low temperature regime, where $\xi > l$, the stripes reduce
the effective superexchange coupling perpendicular to the domain walls and 
cause the pronounced decrease of the spin-stiffness constant. Since no narrowing in  
the $^{139}$La NQR spectra has been observed \cite{bib14}, one can conclude
that the hopping rate for the stripes should be lower than $\simeq 100$ kHz, for $T\leq T_N$. 
In between these two
regimes $\xi_{2D}$ is  of the
order of $l$ and one observes a progressive saturation of $\xi_{2D}$ on decreasing 
temperature (corresponding to 
a reduction of $\rho_s$).  


In conclusion the $^{63}$Cu NQR $1/T_1$ measurements in 
lightly doped La$_{2-x}$Sr$_x$CuO$_4$ evidence that the spin-stiffness constant 
at high temperatures is reduced by doping only to a small extent, consistent with the
idea of formation of stripes. On decreasing temperature 
a progressive saturation for the in-plane correlation length 
is found, with a concomitant decrease in the spin-stiffness. The temperature dependence 
of $\rho_s$ and 
the analysis of the correlation length $\xi_{2D}(x,T)$ support the
hypothesis of  microsegregation of the holes
along stripes.



\section{Acknowledgements}


	The research was carried out with the financial support of
INFM (Istituto Nazionale di 
Fisica della Materia). Ames Laboratory is operated for the U.S. Department 
of Energy by Iowa State University under contract No. W-7405-Eng-82. 
The work at Ames was supported by the Director for Energy Research, 
Office of Basic Energy Sciences.





\begin{figure}
\caption{$^{63}$Cu NQR spin-lattice relaxation rates $2W$  in the paramagnetic 
phase of La$_{2-x}$Sr$_x$CuO$_4$ for different Sr doping levels $x$. 
The lines are guides to the eye.}
\end{figure}


\begin{figure}
\caption{In-plane magnetic correlation length $\xi_{2D}$ 
as a function of inverse temperature in the 
paramagnetic phase of La$_{2-x}$Sr$_x$CuO$_4$, as extracted from $^{63}$Cu NQR spin-lattice 
relaxation rate 
measurements (see text) for a) $x <0.02$ and b) $x >0.02$. The dotted and dashed 
lines are best fits according 
to Eq. (1) with values of $J(x)$ as follows: $J (0.012) = 1340$ K, $J (0.018) = 1200$ K,  
$J (0.024) = 1250$ K,  $J (0.03) = 1160$ K. In a) the solid line shows the corresponding 
behavior of $\xi_{2D}$ for $x=0$, where $J=1588$ K.
In b) the triangles show the data for $x=0.03$ obtained
from neutron scattering  (Ref. 15).}
\end{figure}


\begin{figure}
\caption{Concentration $x$ dependence of the spin-stiffness
 $\rho_s(x)/\rho_s(0)
\propto J(x)/J(0)$ in 
La$_{2-x}$Sr$_x$CuO$_4$ as derived form the fits in Fig.2 (solid circles, 
the dotted line is a guide to the eye). The dashed 
line represents the behavior of $\rho_s(x)/\rho_s(0)$ for a diluted 2D-QHAF, 
while the squares show
the corresponding values derived from $1/T_1$ in Zn-doped La$_{2}$CuO$_4$
(see Ref. 5). The solid
line shows the behavior for $\rho_s(x)/\rho_s(0)$ derived on the basis of the analysis of the
magnetization data (Ref. 14) by Castro Neto and Hone (Ref. 17) (see text).}
\end{figure}



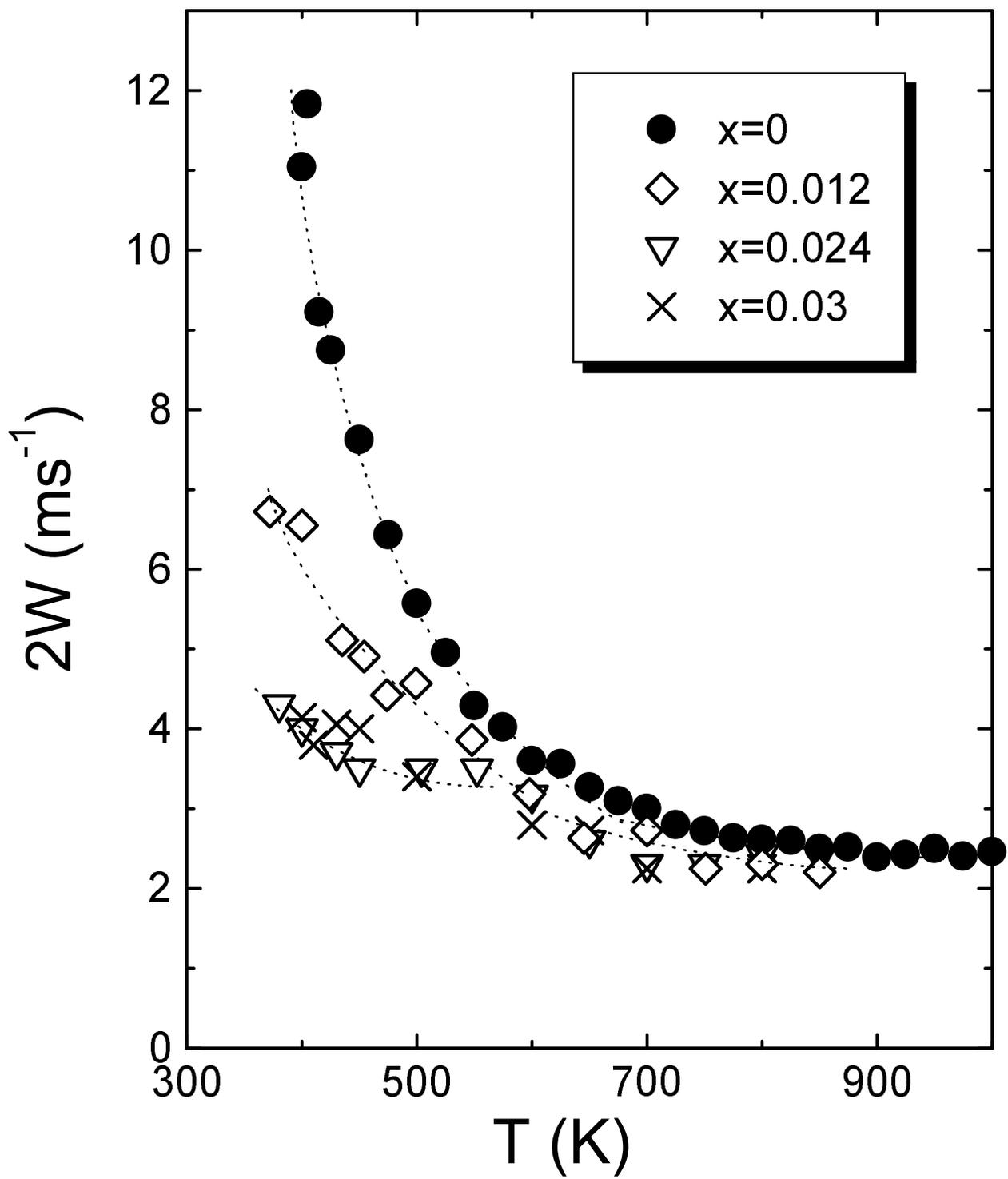
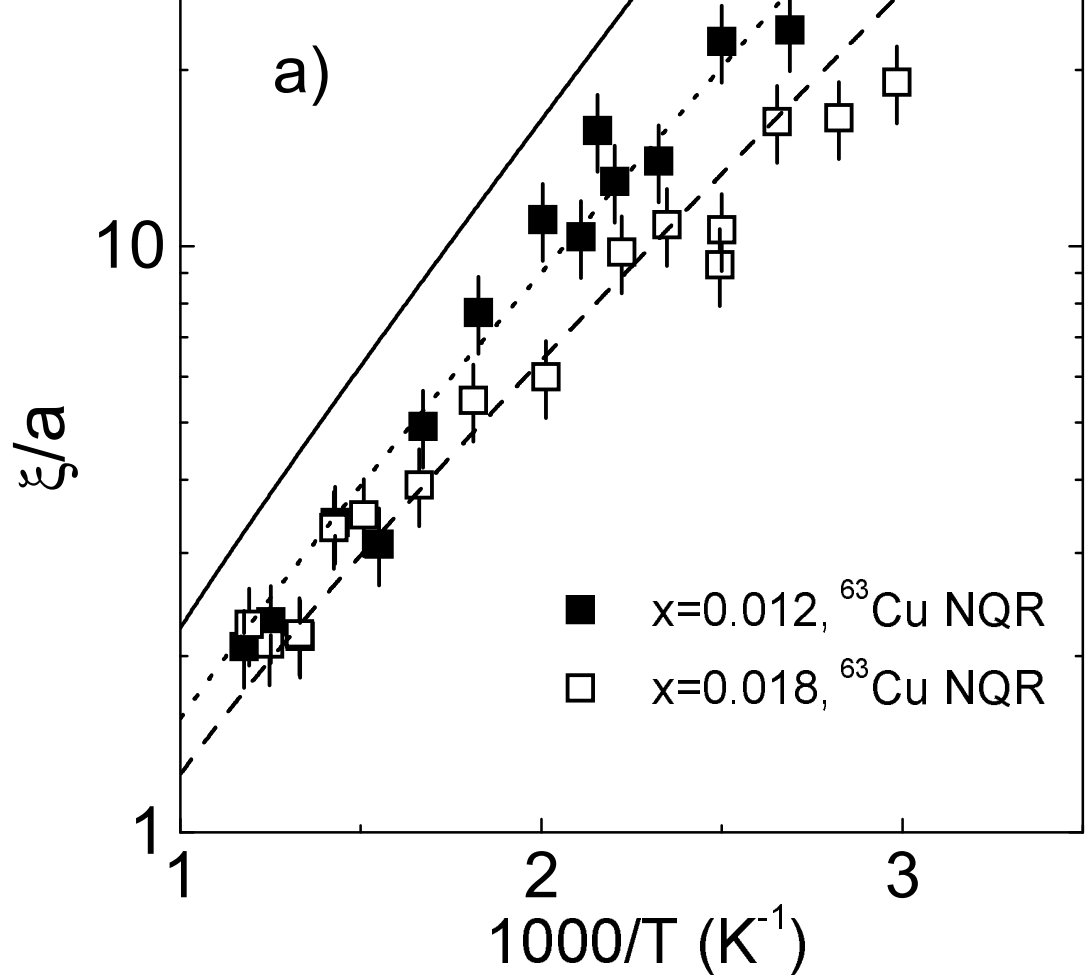
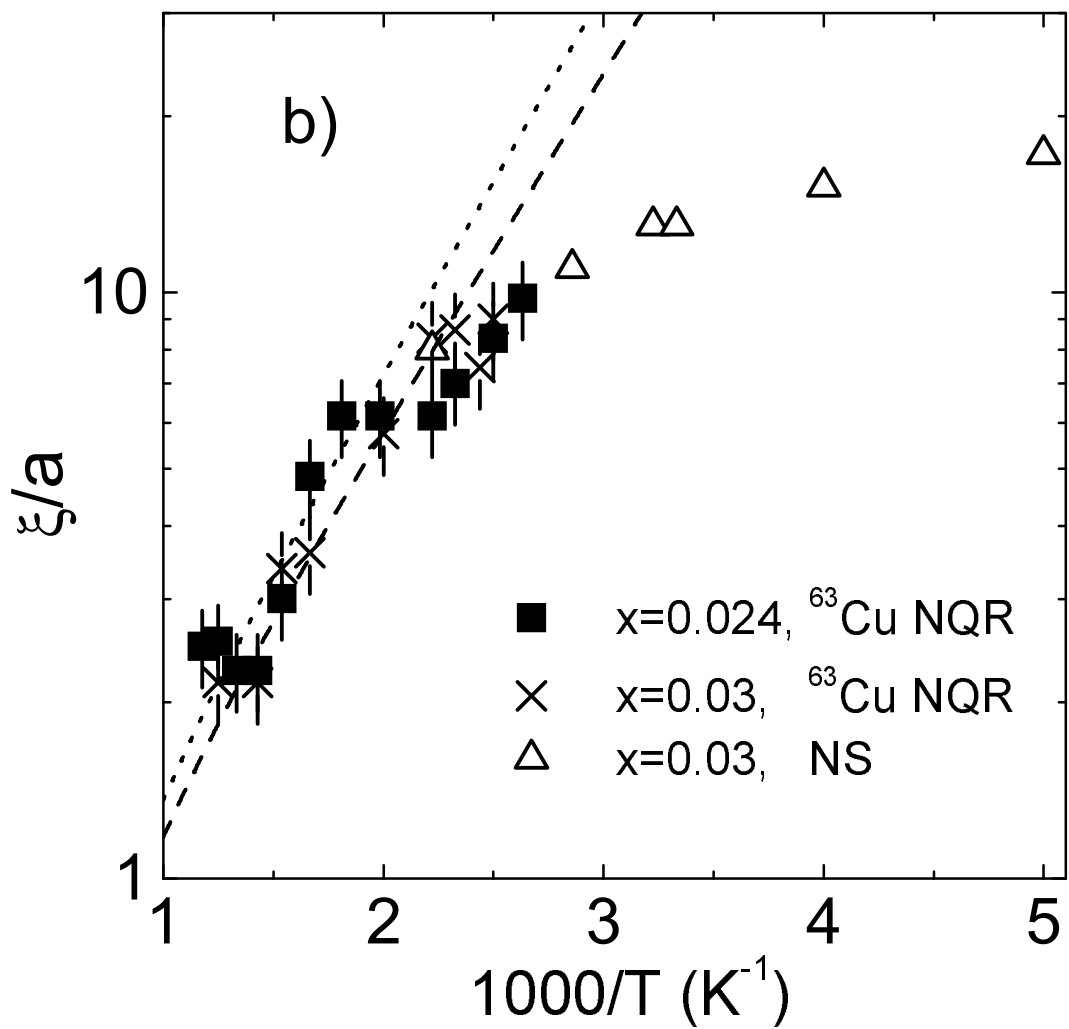
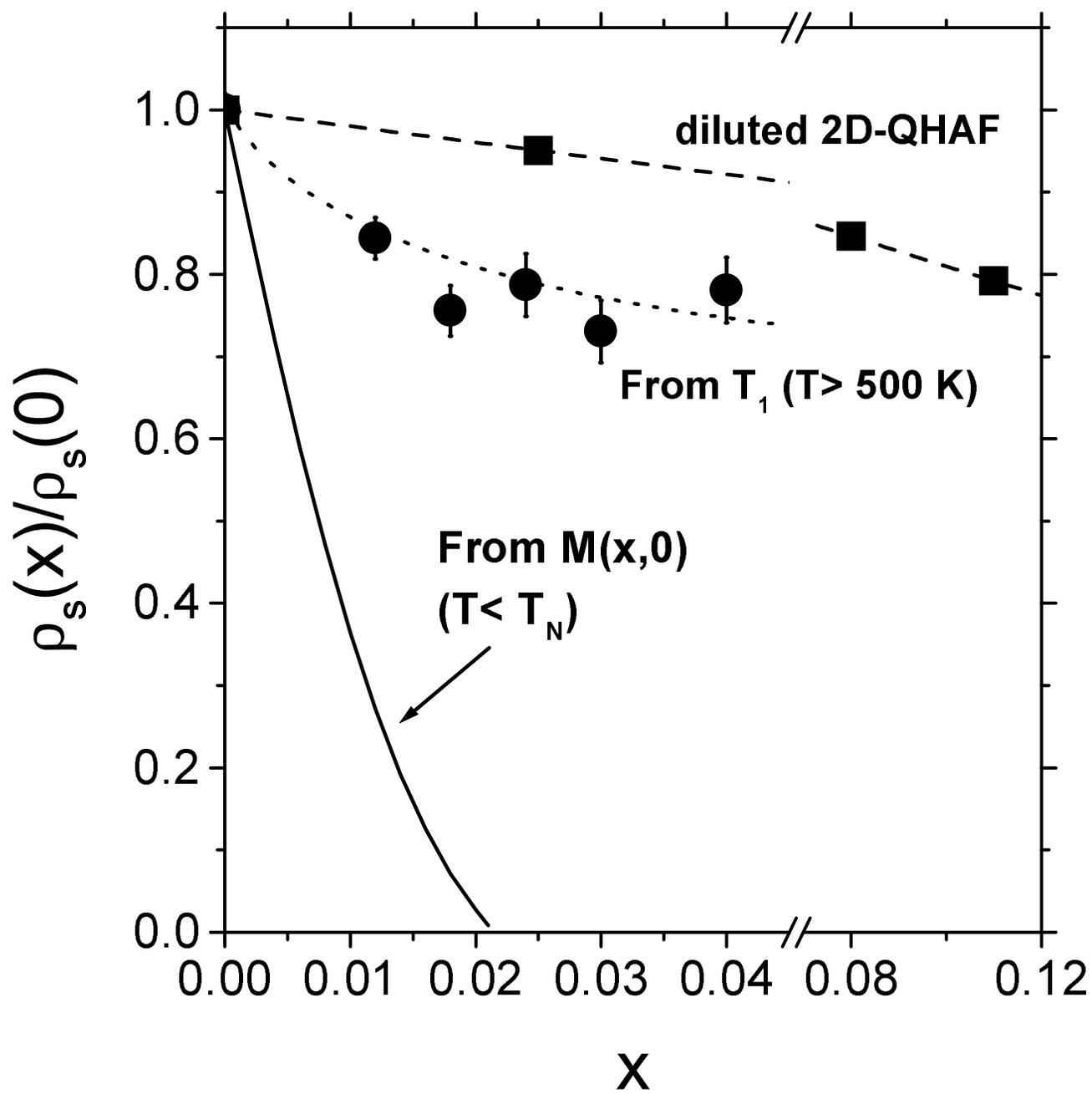
\end{document}